# Comment on "Hartree-Fock –Roothaan Calculations for Ground States of Some Atoms Using Minimal Basis Sets of Integer and Noninteger $n$-STOs"

I.I. Guseinov

*Department of Physics, Faculty of Arts and Sciences, Onsekiz Mart University, Çanakkale, Turkey*

**Abstract**

Recently published formulas for the one-center integrals arising in atomic Hartree-Fock-Roothaan (HFR) calculations with noninteger $n$ STOs (S.Gümüş, T. Özdoğan, Chin. J. Chem., 22 (2004) 1262) are critically analyzed. The purpose of this note is to point out that the presented in these work relations for the integer $n$-nuclear attraction and kinetic energy integrals which are available in the literature (C.C.J. Roothaan, J. Chem. Phys., 19 (1951) 1445) can not be used for noninteger $n$ STOs. In addition, the formulas for two-electron integrals can be obtained from the published in the literature (T. Koga, K. Kanayama, Chem.Phys.Let., 266(1997)123; I.I. Guseinov, B.A. Mamedov, Theor. Chem. Acc.,108 (2002) 21) relations by changing the indices. It should be noted that the accuracy of computer values for ground states energy of some closed and open shell atoms in the case of noninteger $n$ STOs is not guaranteed since the calculations were performed by the use of integer $n$-one electron integrals. It is argued that the paper sheds no new light on the subject and that it is altogether misleading.

**I. Introduction**

It is well known that the noninteger $n$ STOs ( NISTOs) provide a more flexible basis for atomic calculations than usual integer $n$ STOs ( ISTOs) and also the energies calculated by the use of NISTOs lead to better atomic energies than in conventional ISTOs [1,2]. Gümüş and Özdoğan in Ref. [3] published formulas for the one-center one- and two- electron integrals over noninteger $n$ STOs by the use of which performed the HFR calculations for ground states of some atoms. The calculation results of ground states energies (see Tables 3 and 4) in the case of noninteger $n$ STOs were obtained by the use of formulas for the one-center integer $n$- nuclear attraction and kinetic energy integrals (see Eqs.(3) and (4) of Ref.[3]) contained in the literature [4]. The purpose of this Comment is to demonstrate that the relations for one-center two-electron integrals over noninteger $n$ STOs published by Gümüş and Özdoğan in Ref.[3] are not original and they can easily be derived from the relationships given in the literature (see. e.g., Refs.[5, 6]) by changing the indices.

**2. Theory**

The one-center two-electron integrals of noninteger $n$ STOs arising in atomic HFR calculation are as follows:

$$I_{p_1 p_1', p_2 p_2'}(\zeta_1 \zeta_1', \zeta_2 \zeta_2') = \int \chi_{p_1}^*(\zeta_1, \vec{r}_1) \chi_{p_1'}(\zeta_1', \vec{r}_1) \frac{1}{r_{21}} \chi_{p_2}^*(\zeta_2, \vec{r}_2) \chi_{p_2'}(\zeta_2', \vec{r}_2) dV_1 dV_2 , \quad (1)$$

where $p_i \equiv n_i l_i m_i$ and $p_i' \equiv n_i' l_i' m_i'$ $(i=1,2)$. The formulas for multicenter electron-repulsion integrals with noninteger $n$ STOs have been established in Ref.[6]. Using Eqs.(19), (21), (22), (24), (30) and (31) of Ref.[6] it is easy to obtain for the one-center two-electron integrals, Eq.(1), the following relation:

$$I_{p_1 p_1', p_2 p_2'}(\zeta_1 \zeta_1', \zeta_2 \zeta_2') = N_{n_1 n_1'}(t_1) N_{n_2 n_2'}(t_2) \sum_L P_{k_1 k_2}^L(z_1, z_2) \sum_M C^{L|M|}(l_1 m_1, l_1' m_1') A_{m_1 m_1'}^M$$
$$\times C^{L|M|}(l_2 m_2, l_2' m_2') A_{m_2 m_2'}^M , \quad (2)$$

where $\max(|l_1 - l_1'|, |l_2 - l_2'|) \leq L \leq \min(l_1 + l_1', l_2 + l_2')$, $-L \leq M \leq L$, $k_i = n_i + n_i' - 1$, $t_i = (\zeta_i - \zeta_i')/(\zeta_i + \zeta_i')$, $z_i = \zeta_i + \zeta_i'$ and

$$N_{n_i n_i'}(t_i) = \frac{(1+t_i)^{n_i+1/2}(1-t_i)^{n_i'+1/2}}{[\Gamma(2n_i+1)\Gamma(2n_i'+1)]^{1/2}} . \quad (3)$$

The quantity $P_{k_1 k_2}^L(z_1, z_2)$ occurring in Eq.(2) is the radial part of the two-electron integral:

$$P_{k_1 k_2}^L(z_1, z_2) = z_2 \frac{\eta^{k_2+1}}{(1+\eta)^{k_1+k_2+3}} \Gamma(k_1+k_2+3)$$
$$\times \left[ \frac{{}_2F_1(1, k_1+k_2+3, k_1+L+3, 1/(1+\eta))}{k_1+L+2} + \frac{{}_2F_1(1, k_1+k_2+3, k_2+L+3, \eta/(1+\eta))}{k_2+L+2} \right] , \quad (4)$$

where $\eta = z_2/z_1$ and ${}_2F_1(a,b,c;x) \equiv F(a,b,c;x)$ is the Hypergeometric function. The relation (4) for $P_{k_1 k_2}^L(z_1, z_2)$ is also available in Ref.[5] (see Eq.(3) of Ref.[5]). It is easy to show that the Eqs.(5)-(9) occurring in Ref.[3] can be obtained from Eqs.(1)-(4) of this work by changing the indices. Thus, all of the formulas given by Gümüş and Özdoğan are not original and they are available in Refs.[4-6].